\documentclass[referee]{aa}
\usepackage{graphicx}
\begin{document}
\title{Compressible 2D MHD modelling of prominence dips}

\titlerunning{MHD Prominence Models}
\authorrunning{Petrie et al.}

\author{\bf G.J.D.~Petrie
          \inst{1}\fnmsep\thanks{Present address:
             High Altitude Observatory
             National Center for Atmospheric Research
             PO Box 3000, Boulder, CO 80307-3000, USA}
          \,
          K. Tsinganos
          \inst{1}
          \,
          T. Neukirch
          \inst{2}
          }

   \offprints{G.J.D. Petrie}

   \institute{IASA and Section of Astrophysics, Astronomy and Mechanics,
Department
of Physics, University of Athens, Panepistimiopolis, GR-157 84 Zografos,
Athens, Greece
              \email{gordonp@phys.uoa.gr,tsingan@phys.uoa.gr}
         \and
              School of Mathematical and Computational Sciences,
University of St. Andrews, St. Andrews, KY16 9SS, Scotland\\
              \email{thomas@mcs.st-and.ac.uk}
             }

   \date{Received;accepted}

\abstract{
An analytical MHD model of a normal-polarity prominence with compressible
flow is presented. The exact solution is constructed via a systematic 
nonlinear separation of variables method used to calculate several classes 
of MHD equilibria in Cartesian geometry and uniform gravity.
Although the model is 2D, a third magnetic/velocity vector field component
is included and the highly sheared fields observed in prominences are
reproduced.  
A description is given of the balance of gas pressure gradient with gravity 
and the Lorentz or inertial forces acting along and across the prominence.  
It is found that the flow does not significantly influence the heating
profile.  The analyzed model has dimensions, plasma density, temperature and
velocity profiles which agree with those in the observations literature. 
}

   \maketitle
%

\section{Introduction}
\label{introduction}

The term {\sl prominence} is used to describe various objects ranging from relatively
stable ones with lifetimes of many months to transient phenomena lasting hours
or less (Tandberg-Hanssen,~1995).  When they are seen in absorption against the
disk they are referred to as {\sl filaments}.  The long-lasting structures observed to
last from days to months away from active regions are often called {\sl quiescent
prominences}.  They are long, cool, dense, sheet-like structures near-vertical
to the solar surface supported by a series of arches whose feet are anchored
in the photosphere.  In and around active regions, a different kind of
shorter-lived prominences exist, refered to as {\sl active region
prominences}.  On the disk their appearance is like that of quiescent
prominences except that they are generally smaller.  The category of active
region prominences can be further subdivided into plage filaments, which are
relatively stable prominences found above magnetic polarity inversion lines in
or bordering active regions, and more dynamic phenomena such as surges, sprays
and flare loops (Tandberg-Hanssen,~1995).

Quiescent prominences are structures of cool plasma suspended in the
chromosphere or corona, usually above photospheric polarity inversion
lines.  A prominence is said to be of normal or inverse polarity depending
on whether the prominence field points in the same direction across the
prominence as the field of the bipolar region below, or in the opposite
direction.   
Before the crucial role played by magnetic fields in prominence physics was
understood, prominences were regarded as cool objects in hydrostatic equilibrium
with the hot corona.  However, Menzel (Bhatnagar et al.,~1951) argued that
coronal magnetic fields could support prominences in static equilibrium.  
Different formulations of this problem were given by Dungey~(1953), Kippenhahn
\& Schl\"uter~(1957) and Brown~(1958).  In these models the dense prominence
material is supported against gravity mainly by the Lorentz force.  Since then
many models of prominence support have been developed, most of them
two-dimensional because prominences are observed to be long, straight and
reasonably uniform along their long axes structures.  Such normal and inverse polarity
models include those by Anzer~(1972), Kuperus \& Raadu~(1974), Lerche \& Low~(1977),
Malherbe \& Priest~(1983), Anzer \& Priest~(1985), Hood \& Anzer~(1990),
Fiedler \& Hood~(1992), Low \& Hundhausen~(1995), Low \& Zhang~(2002), Fong
et al.~(2002) and Low et al.~(2003).

Because of the mathematical complexity of the full 3D magnetohydrostatic
equations most prominence modelling is 2D.  However, important exact 3D
magnetostatic prominence models have been calculated by Low~(1982, 1984, 1992).  
Since the full 3D MHD equations are not amenable to analytical treatment (but see
Petrie \& Neukirch, 1999) we will focus on the basic macroscopic structure of
a prominence. Full MHD normal prominence models have rarely been attempted 
in the past (in 1D by Tsinganos \& Surlantzis,~1992; in 2D by Ribes \& Unno, 
~1980; Del Zanna \& Hood,~1996).

As well as support against gravity, also important is the energy balance within
a prominence.  The energy balance in prominences has been modelled using
radiative transfer theory by many authors, e.g. Poland et al.,~(1971), Heasley
\& Mihalas~(1976), Heinzel et al.,~(1987), Paletou et al.,~(1993), Gontikakis et 
al.,~(1997) and Anzer \& Heinzel~(1999, 2000) while Poland \& Mariska~(1986) have
modelled 1D normal prominences with a unidirectional flow and asymmetric heating.  
In this project we include a full MHD momentum balance and a simple treatment 
of the energy balance in a prominence model for the first time.
The focal object of the present study is to investigate the effect of 
non-isothermal compressible flows in prominence dips for the first time and, 
following on from Paper 2, to check if these flows influence the heating as 
such flows do in coronal loops.  

The paper is organised as follows.  The analytical modelling technique is
outlined in Sect. \ref{analyticalmodel}.  A model fitted to
typical observed physical parameter values is presented in Sect.
\ref{models} and the results are summarized in Sect. \ref{conclusions}.

\section{The analytical model}
\label{analyticalmodel}

This model uses the first family in Table~1 of Petrie et al.~(2002,
henceforth Paper 1) generalising the models by
Hood \& Anzer~(1990) and Del Zanna \& Hood~(1996), but not the models by
Kippenhahn \& Schl\"uter (1957) and Tsinganos \& Surlantzis~(1992) which are
from the second family.  For the first time, however, our solutions are 
non-adiabatic and non-isothermal so that a study of the effect of the flow on the
thermodynamics of our solutions will be possible as in Petrie et al.~(2003,
henceforth Paper 2).

The dynamics of flows in solar coronal loops may be
described to zeroth order by the well known set of steady
(\(\partial/\partial t=0\)) ideal hydromagnetic equations:
\begin{equation}\label{momentum}
\rho \left( {\bf V}\cdot{\bf\nabla}\right){\bf V}= \frac{1}{4 \pi}
{\left({\bf\nabla}\times{\bf B}\right)\times{\bf B} }
-{\bf\nabla}P-\rho g {\bf\hat Z} \,,
\end{equation}
\begin{equation}\label{fluxes}
\bf{\nabla}\cdot{\bf B}=0\,,\quad \bf{\nabla}\cdot\left(\rho{\bf
V}\right)=0 \,,\quad \bf{\nabla}\times\left({\bf V}\times{\bf
B}\right)=0 \,,
\end{equation}

\noindent where \({\bf B}\), \({\bf V}\), \(-g {\bf\hat Z} \)  denote
the magnetic, velocity and (uniform) external gravity fields while
\( \rho\) and $P$ are the gas density and pressure.
At present, a fully three-dimensional MHD equilibrium model with
compressible flows is not amenable to analytical treatment and so we
assume translational symmetry. Thus, we assume that in Cartesian
coordinates $(Z, X, Y)$, the
coordinate $Y$ is ignorable (\(\partial/\partial Y=0\)).  We are, however, able
to include a $Y$-component in the model and thereby model magnetic shear. 
Meanwhile the
energetics of the flow are governed by the first
law of thermodynamics :
\begin{equation}\label{firstlaw}
q=\rho {\bf V} \cdot \left [ {\bf\nabla} e+P {\bf\nabla} \frac{1}{\rho} \right
]
=\rho {\bf V} \cdot \left [ {\bf\nabla} h-\frac{1}{\rho}{\bf\nabla} P \right ]
\,, \label{qdef}
\end{equation}
where $q$ is the net volumetric rate of some energy input/output,
$\Gamma=c_{p}/c_{v}$ with $c_{p}$ and $c_{v}$ the specific
heats for an ideal gas, and

\begin{equation}
e=\frac{1}{\Gamma-1}
\frac{P}{\rho}
\end{equation}

\noindent the internal energy per unit mass, with $h=\Gamma e$ the
corresponding  enthalpy function.  For a detailed description of the 
solution method see Papers 1 and 2.

Finally, consider the energy balance along the loop; the net volumetric rate 
of heating input/output $q$, equals to the sum of the net radiation $L_R$, the
heat
conduction energy $\nabla\cdot{\bf F}_C$, where ${\bf F}_C$ is the
heat flux due to conduction, and the (unknown) remaining heating
$E_H$,

\begin{equation}
q=E_H+L_R-\nabla\cdot{\bf F}_C.
\end{equation}

\noindent The net heat in/out $q$ is calculated from the MHD model using the
first law of thermodynamics Eq. (\ref{firstlaw}).  If the ionisation is
dominated by collisional processes as in the corona then the assumption of
local thermodynamic equilibrium (LTE) is valid.  This may not hold for prominences,
which consist of thin flux tubes of width
approximately 300-1000~km (D\'emoulin et al.,~1987) which
are irradiated from the much hotter corona as well as the transition region
and the chromosphere (Tandberg-Hanssen,~1995).  The ratio of
radiatively-induced transitions to collisionally-induced transitions may be
large and the plasma far removed from LTE.  The radiative losses in
prominences are dominated by hydrogen (Zhang \& Fang,~1987).  In very tenuous
plasmas such as in the corona the hydrogen spectral lines may be considered to
be optically thin. 
However in most prominences the hydrogen and helium lines become optically
thick at $40,000$~K and the optically thin approximation overestimates the
losses (Kuin \& Poland,~1991).
In our model we include non-LTE effects by incorporating the radiation model
of Kuin \& Poland~(1991).  Inside the prominence, therefore, the radiative
losses are given by

\begin{equation}
L_R=-N(e)N(H)\phi (p, T),
\end{equation}

\noindent where $N(e)$ is the electron number density, $N(H)$ is the number 
density of hydrogen ions or atoms and the function $\phi (p, T)$ is tabulated in Kuin \&
Poland~(1991), Table~1.  The thermal conduction
energy is calculated assuming that conduction is mainly along the field, using
the expression

\begin{equation}
-\nabla\cdot{\bf F}_C =  \frac{\partial}{\partial s}\left(\kappa_{\small
||}\frac{\partial T}{\partial s}\right) -\frac{\kappa_{\small
||}}{B}\frac{\partial B}{\partial s}\frac{\partial T}{\partial s}
,\label{conduction}
\end{equation}

\noindent (Spitzer, 1962) where subscripts $||$ indicate values and derivatives
along the
field line, and the variation of the magnetic field strength along the field
line is taken into
account (Priest, 1982, p86).

Including a $Y$-component in the model the expressions for the physical
quantities are (compare with Paper 2, Eqs.~(17-22))

\begin{eqnarray}
A(\alpha ) & = & Z_0 B_0 \int\sqrt{2C_1 +\lambda C_2 \alpha^{\lambda
-2}}d\alpha \,,\qquad \alpha=G(x) \exp{(-z)}\,\label{G}\\
\Psi_A (\alpha ) & = & \frac{B_0}{\sqrt{gZ_0}} \sqrt{2D_1 \alpha^2
+\lambda
D_2\alpha^{\lambda}} ,\\
\rho (x,\alpha )& = & \frac{B_0^2}{4 \pi g Z_0} \frac{2D_1 \alpha^2
+\lambda
D_2\alpha^{\lambda}}{M^2} ,\\
P(x,\alpha ) & = & \frac{B_0^2}{4\pi} \left[ P_0+P_1(x) \alpha^2+ P_2(x)
\alpha^{\lambda}\right] -\frac{B_0^2 L^2}{8\pi g
Z_0} \frac{2D_1 \alpha^2 +\lambda D_2\alpha^{\lambda}}{(1-M^2)^2} 
,\label{pressuresum}\\
{\vec B} & =& B_0 \sqrt{2C_1 \alpha^2 +\lambda C_2\alpha^\lambda}
\left[{\bf\hat{X}}+F(x) {\bf\hat{Z}}\right] -\frac{B_0 L}{\sqrt{g Z_0}}
\frac{\sqrt{2D_1 \alpha^2 +\lambda D_2\alpha^{\lambda}}}{1-M^2}
{\bf\hat{Y}},\\
{\vec V} & =& \sqrt{g Z_0} \sqrt{\frac{2C_1 \alpha^2 +\lambda
C_2\alpha^{\lambda}}{2D_1 \alpha^2 +\lambda D_2\alpha^{\lambda}}} M^2
\left[{\bf\hat{X}}+F(x) {\bf\hat{Z}}\right] -\frac{L
M^2}{1-M^2} {\bf\hat{Y}}.
\end{eqnarray}

\noindent where $C_1$, $C_2$, $D_1$,  $D_2$ and $\lambda$ are constants and 
$\Psi_A(\alpha )$ is the mass flux per unit of magnetic flux.  Note
that although these solutions are invariant in $Y$ the inclusion of a
$Y$-component in the model allows us to model the sheared magnetic and velocity
fields which characterise prominences.

In the expression for the pressure $P_0 = f_0 =$ constant, while 
$P_1$ and $P_2$ satisfy the following two ODE's

\begin{eqnarray}
P_1&=&C_1 \left[FM^{2'}- F^{'} (1-M^2)-F^2-1\right]+\frac{D_1}{M^2} \,,
\nonumber \\
P_2&=&C_2 \left[FM^{2'}- F^{'}
(1-M^2)-\frac{\lambda}{2}(F^2+1)\right]+\frac{D_2}{M^2} \,.\nonumber
\end{eqnarray}

\noindent Using the above definitions for the pressure ``components'' together
with the ODE's from Table 1 in Paper 1, we calculate that for
the general case we have the following final system of equations for the
unknown functions of $x$, including the slope of the field lines $F$ :

\begin{eqnarray}
\frac{d \ln G}{dx} & = & F\,, \label{Gd}\\
M^{2'} (x)& = & \frac{\mathcal{C}\lambda F/M^2 - 2F(F^2
+1+P_1/C_1)}{\mathcal{C}/M^4
+ 2}\,,\ \mbox{where}\ 
\ \mathcal{C} = \frac{D_2/C_2-D_1/C_1}{1-\lambda /2}\,, \label{M2d}\\
F ' (x)& = & \frac{FM^{2'} - F^2 -1 + D_1/C_1 M^2 -P_1/C_1}{1-M^2} \,,
\label{Fd}\\
P_1 ' & = & -\frac{2 D_1 F}{M^2} - 2 C_1 (1+F^2 ) M^{2'} - 2 M^2 FF'
\,.\label{P1d}\\
P_2 (x) & = & C_2 \left( F M^{2'} -F'(1-M^2 ) - \frac{\lambda}{2} ( 1+F^2 )
\right)
+\frac{D_2}{M^2 }\,, \label{P2}
\end{eqnarray}
where
\begin{equation}
\mathcal{C} = \frac{D_2/C_2 - D_1/C_1}{1-\lambda /2}
\,.
\end{equation}

We integrate this set of ODEs to get a complete solution.

\section{Results}
\label{results}
In the following sections we present the results of the integretion of the 
previous ODEs in constructing a prominence model, discuss the composition 
of the plasma and the applicability of ideal MHD and finally present 
and discuss a breakdown of momentum and energy balance along and across the 
prominence.  

\subsection{The models}
\label{models}

Brown~(1958) was able to describe analytically the distinction between the
periodic and non-periodic cases of the solutions by Menzel
(Bhatnagar et al.,~1951). 
For these one-dimensional solutions the second-order ODE for $G(x)$ may be
integrated to give solution curves in the $G(x),F(x)=G'(x)$ phase space. 
Closed phase curves correspond to periodic solutions and open curves to
non-periodic solutions.  In our more general case such an explicit description
is not possible.  However it can be seen that if part of our expression for
$F'(x)$ has a single sign and dominates the expression, then $F'(x)$ does not
change sign and the solution is non-periodic.  This depends on the relative
size of the functions $C_1 (F^2+1)$, $D_1/M^2$ and $P_1$.  If any of these
quantities dominates in Eq.~(\ref{Fd}) for $F'(x)$ then we have a single
loop/dip. If two or three of these quantities are comparable then we may have
periodically fluctuating parameters and a periodic solution.  In our
prominence case the dominant quantity is $D_1/M^2$, associated with the
very large density.  This causes $F'(x)$ to be strictly positive and we have a
dip, a field line which is concave upwards.  A large decrease in density and
increase in temperature would result in a dominant pressure function, strictly
negative $F'(x)$ and a loop/arcade solution as in Papers 1 and 2.  If,
however, $C_1 (F^2+1)$ is much larger than $D_1/M^2$ and $P_1$ we have a
flow-dominated solution and negative $F'(x)$.  An arcade solution resembling
the shape of a projectile's trajectory in uniform gravity results as pointed
out by Tsinganos \& Surlantzis~(1992).  The Kippenhahn \& Schl\"uter~(1957)
case is the one where the density's influence is much larger than that
of the flow, i.e. $C_1 (F^2+1)/(D_1/M^2) < 1$ so that $D_1/M^2$ dominates,
$F'(x)$ is positive and we have a dip. 

\begin{figure*}
\begin{center}
\resizebox{0.7\hsize}{!}{\includegraphics*{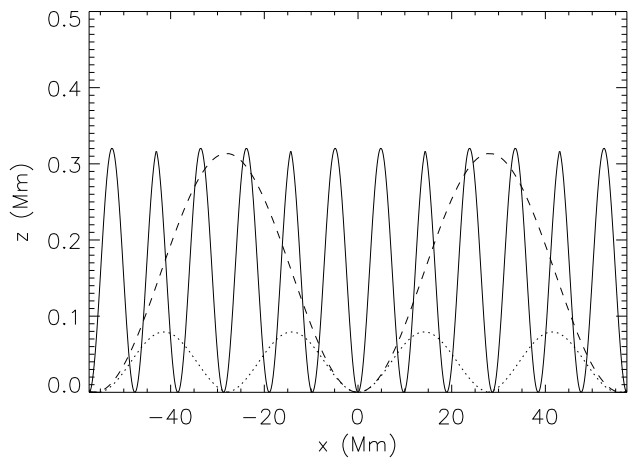}}
\end{center}
\caption{Selection of periodic solutions.}
\label{perline}
\end{figure*}

Fig. \ref{perline} shows a selection of periodic solutions.  The solid line is
a solution whose temperature at the bottom of the dip is $8000$~K, whose
velocity is $10$~km/s, magnetic field strength 2~G and gas density
$3.51\times 10^{-13}$~g/cm$^3$.  It is clearly periodic with period about
10~Mm.  The dotted line is a solution the same as the solid line except that
the magnetic field strength is only 1~G.  The dashed line is a solution the
same as the solid line except that the velocity is 20~km/s.  Both the fast
solution and the solution with weak magnetic field have larger periods than
the first solution because the fluid does not change direction so much, either
because the inertial force is large in the case of the fast solution or
because the inertial force has less effective opposition in the case of the
solution with weak magnetic field.  The solution with weak magnetic field has
a smaller amplitude than the other two because the weaker magnetic field does
not deviate the fluid as far from a straight path.

Of course prominence structure is not periodic.  We discuss now the models we
calculate to reproduce observed features of prominences. 
From measurements of all four Stokes parameters and the Hanle effect the
magnetic vector field can be determined.  A representative value for the field
strength is 8~G for quiescent prominences and 20~G for active region filaments
(Leroy et al.,~1984).  The magnetic field of the
prominence is generally far from perpendicular to the prominence sheet.  The
angle between the magnetic field and the prominence long axis is quite small
and the magnetic field is highly sheared.  Leroy et al.~(1984) derive
statistical results from simultaneous observations of a sample of 256
prominences of low to medium latitude in two optically-thin lines.  They find
that prominences lower than $30,000$~km high are generally of normal
configuration and have shear angle (angle between field lines and prominence
long axis) of average $20^{\circ}$ while those higher than $30,000$~km are of
inverse configuration and have average shear angle $25^{\circ}$.  Computing
the complete 3D structure of a smaller sample of prominences
Bommier et al.~(1994) calculated that the average angle between the
outgoing prominence field and the solar surface is $0^{\circ}$.

\begin{figure*}
\begin{center}
\resizebox{0.49\hsize}{!}{\includegraphics*{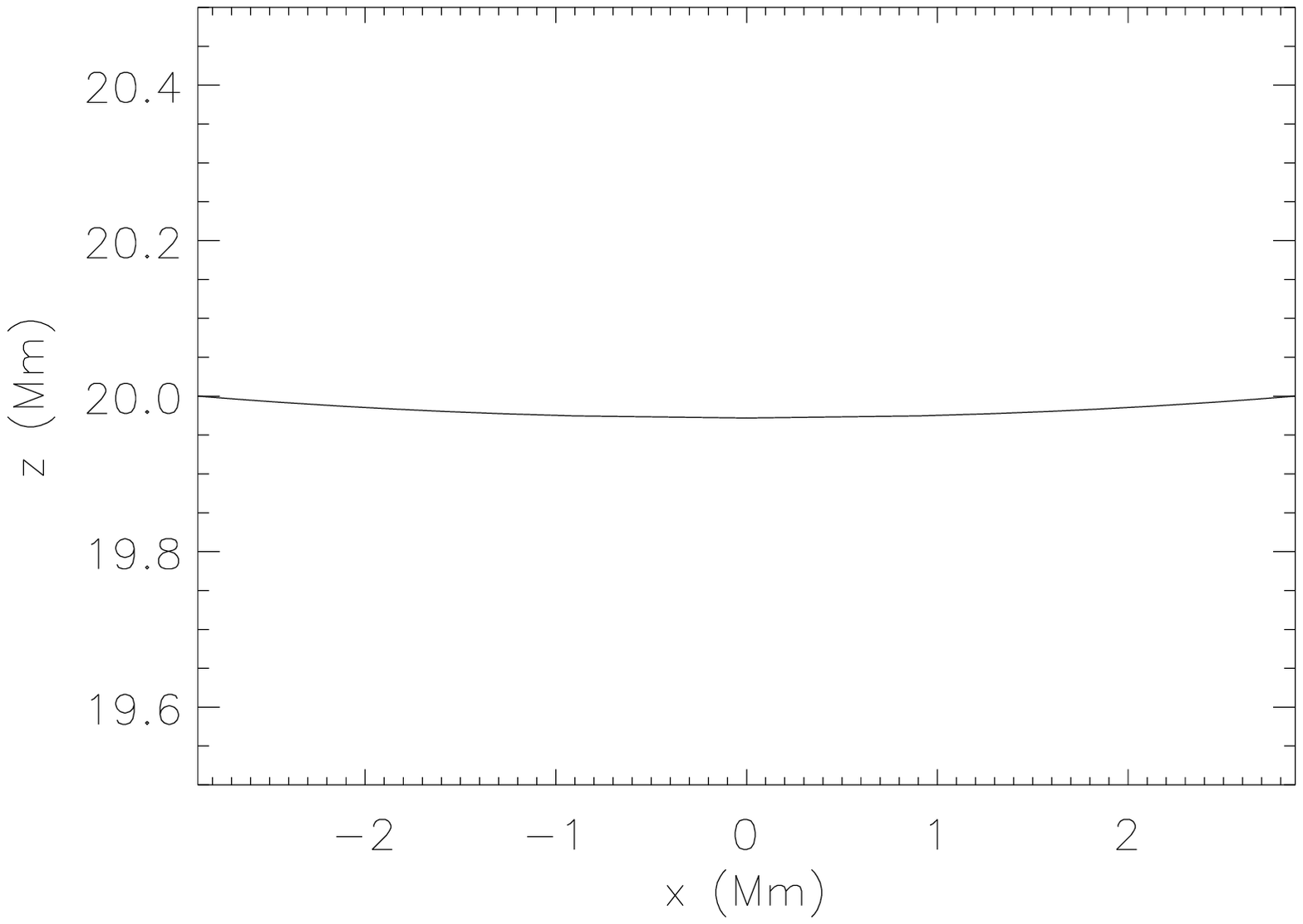}}
\resizebox{0.49\hsize}{!}{\includegraphics*{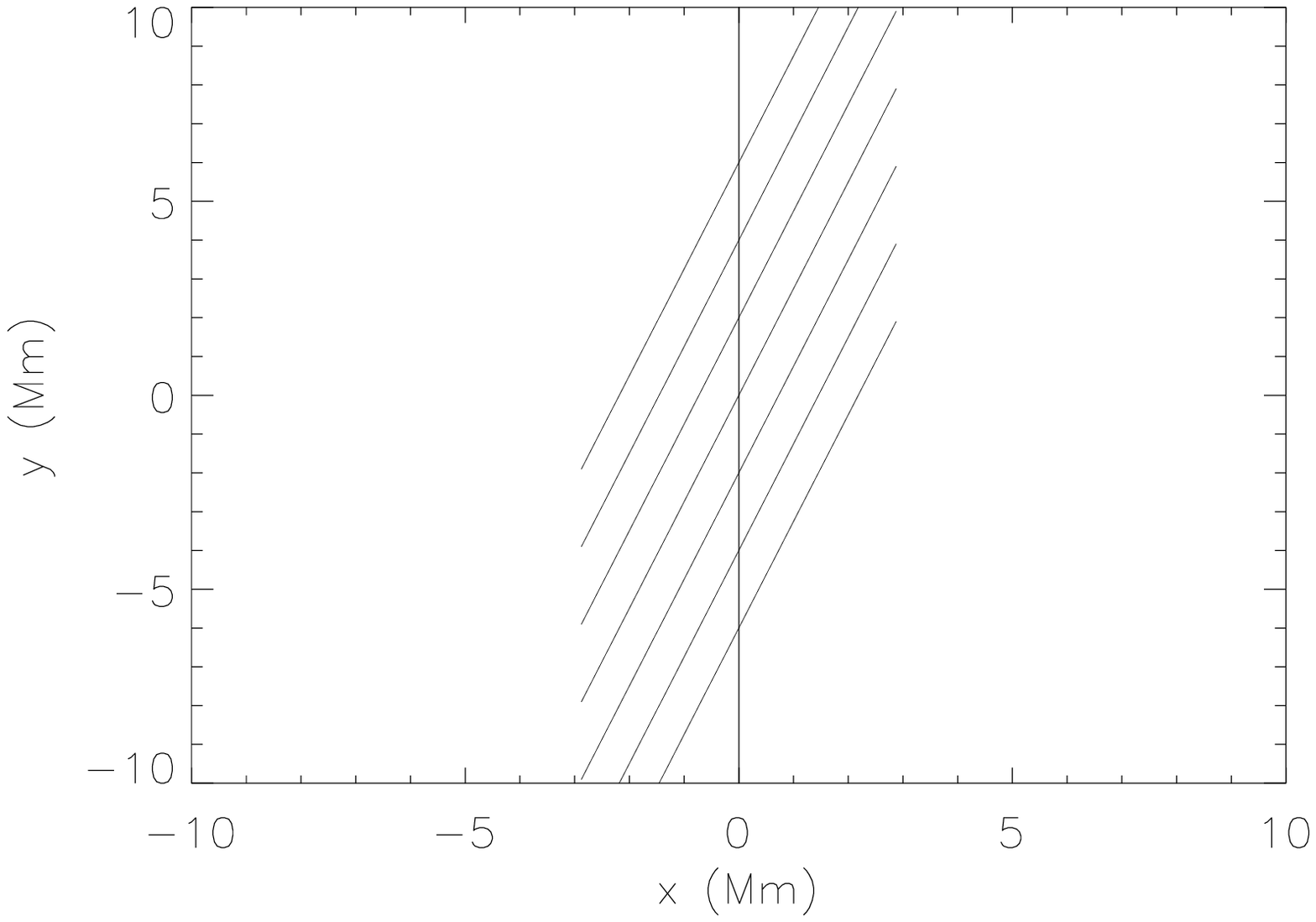}}
\end{center}
\caption{Prominence field line projection in $X$-$Z$ plane (left) and $X$-$Y$ plane (right) with 
the prominence dip and polarity inversion line at $X=0$.}
\label{fieldlinepics}
\end{figure*}

The field line plots of our example model are shown in
Fig.~\ref{fieldlinepics}.  The $Y$-axis points in the direction of the
prominence long axis, the $X$-axis is in the horizontal direction perpendicular
to the $Y$-axis and the $Z$-axis points vertically upwards.  The left
picture shows a plot of the magnetic field line of the prominence projected
onto the $X$-$Z$ plane from the bottom of the prominence dip at $X=0$~Mm to
the edge of the prominence at about $X=2.9$~Mm.  The right picture shows a
selection of field lines projected onto the $X$-$Y$ plane showing the highly
sheared structure of these field lines.  The model is invariant in the $Y$
direction.  The prominence polarity inversion line is also shown.
The prominence width is between $5000$ and
$6000$~km, the height is around $20, 000$~km and the angle between the
prominence field and the long axis is $20^{\circ}$.  The prominence field
is slightly dipped but is nearly flat.  The outgoing field makes an angle
with the solar surface of approximately $0^{\circ}$.

\subsection{Composition of prominence plasma}
\label{composition}

\begin{figure*}
\begin{center}
\resizebox{0.49\hsize}{!}{\includegraphics*{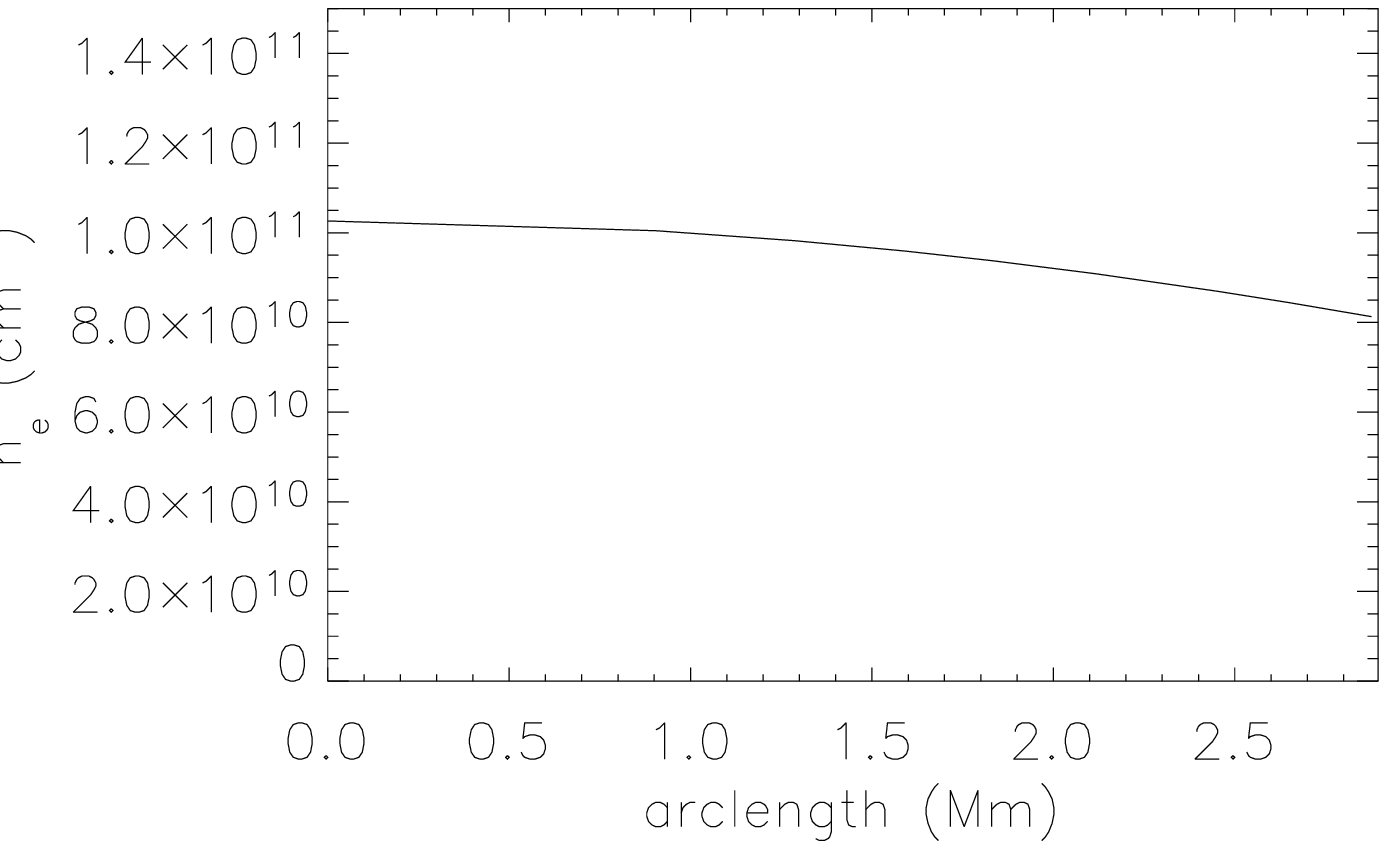}}
\resizebox{0.49\hsize}{!}{\includegraphics*{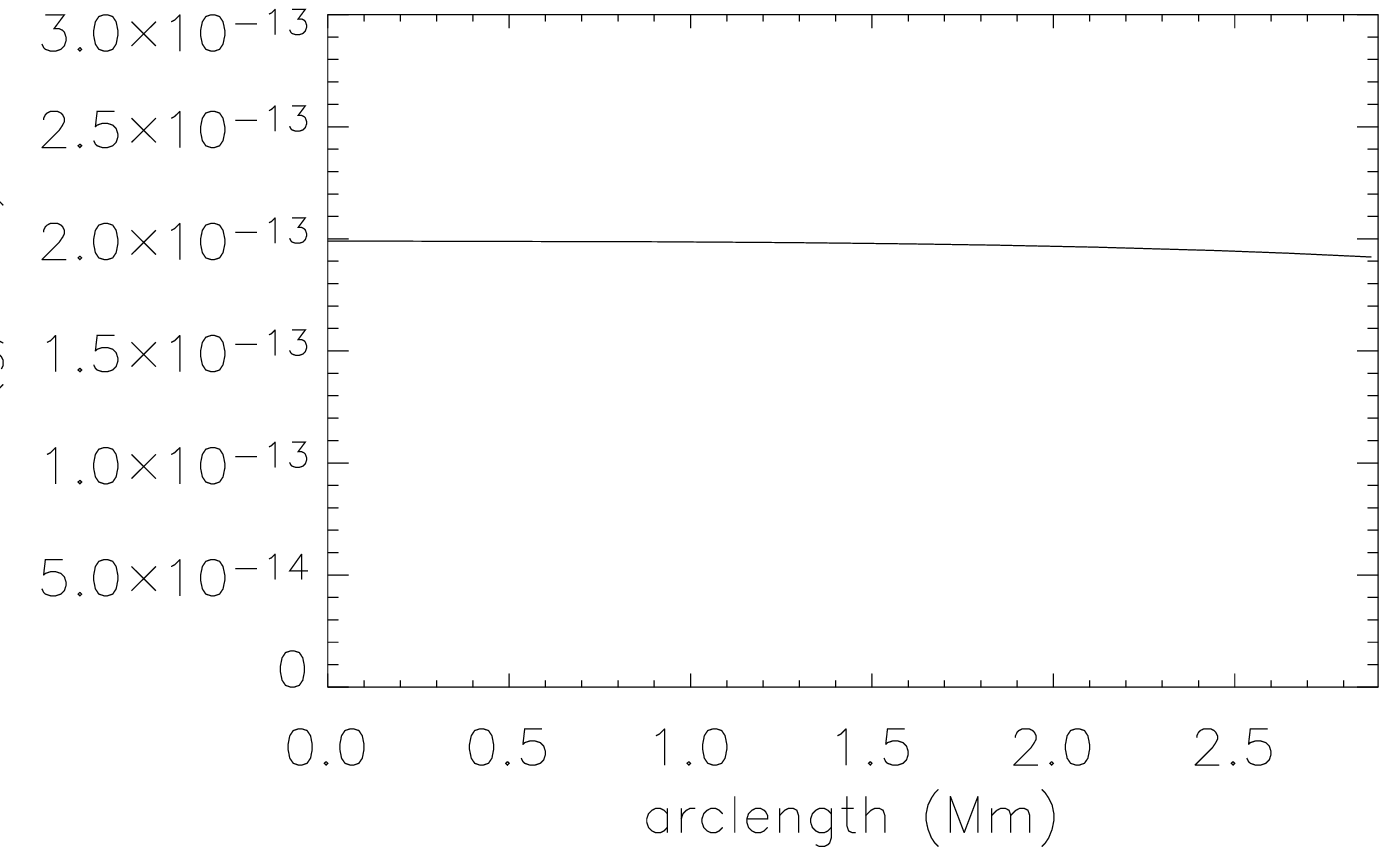}}
\resizebox{0.49\hsize}{!}{\includegraphics*{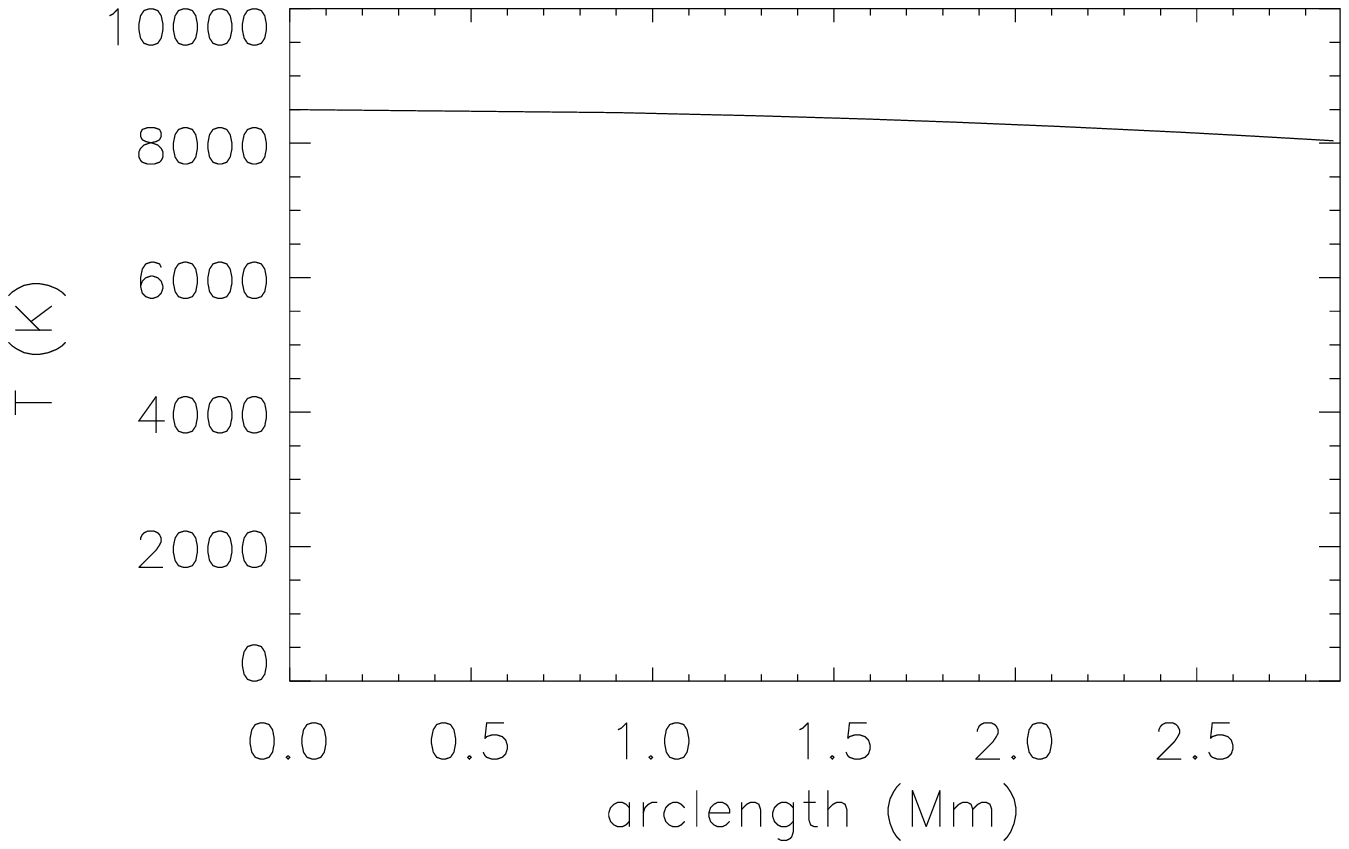}}
\resizebox{0.49\hsize}{!}{\includegraphics*{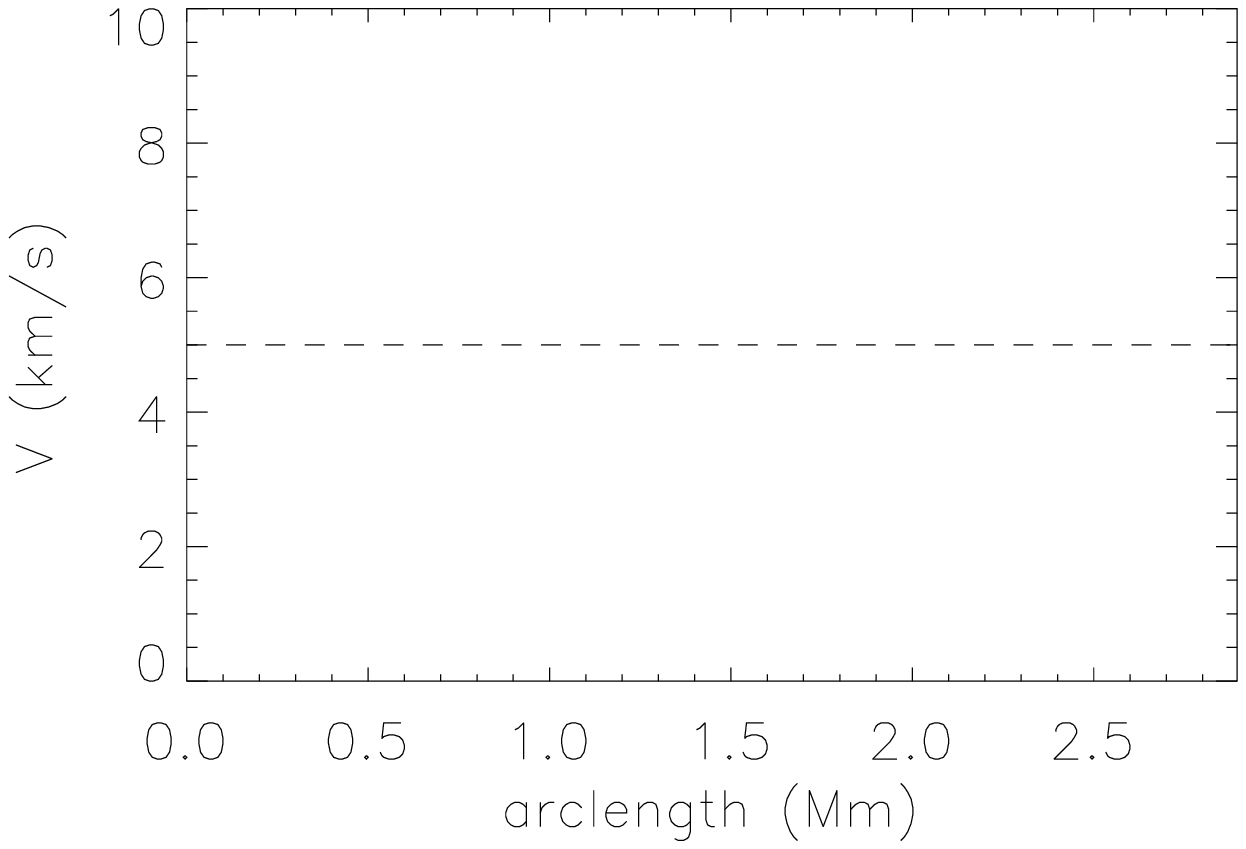}}
\end{center}
\caption{Number density (top left), gas density (top right), temperature
(bottom left) and velocity (bottom right) plots for inside the prominence.}
\label{parampics}
\end{figure*}

Fig.~\ref{parampics} shows the electron number density (top left), the
gas density (top right), the temperature (bottom left) and the flow velocity
(bottom right) inside the prominence,
all graphed against the arclength of the field line projected in the $X$-$Z$
plane.
Typical values for the electron temperature are $5000$~K-$8000$~K
(Hirayama,~1985; Engvold \& Brynildsen,~1986).  Prominence electron number
densities are calculated to be about $10^{11}$-$10^{12}$~cm$^{-3}$
(Hirayama,~1985; Bommier et al.~1986; Landman,~1986; Wiik et al.~1992, 1993). 
The material in quiescent prominences has been observed in movies to be
concentrated in near-vertical thin ropes of diameter less than 300~km (Dunn,
1960; Engvold, 1976).  However the classical picture of flow up or down
vertical ropes (Engvold,~1976) does not agree with the arcade-like structure
of prominences.  When Doppler shifts are measured (Mein, 1977; Martres et al.,
1981; Malherbe et al.,1983; Simon et al., 1986; Zirker et al.,~1993) both blue
and red shifts are found in long loop-like flux tubes possibly connecting
arcade foot-points (Tandberg-Hanssen,~1995).  Mein~(1977) measured H$\alpha$
Doppler shifts in an active region filament on the disk and found zero
vertical velocity on the filament and increasing velocity towards both edges
with opposite sign from each other, a pattern suggesting velocity loops
inclined at small angles to the long axis of the filament.  The observed 
velocity had a local maximum of 5.5 km/s.  Vial et al.~(1979)
found consistent results and drew the same conclusions for an active region
prominence observed in Mg II.

Because the prominence plasma is very dense and because the model is
non-isothermal and compressible, imposing a significant dip angle on the
prominence magnetic field vector causes the plasma at the bottom to become
greatly compressed and unrealistically hot.  This is because there is so much
mass trapped in the dip and even a small slope in the field line can cause a
large pressure gradient.  We were able to produce reasonable temperature
profiles for the prominence model only by imposing a very small dip angle such
as in the configuration shown in Fig.~\ref{fieldlinepics}.  This is in
agreement with the average dip angle for normal prominences observed by Leroy
et al.~(1984) which is approximately $0^{\circ}$.
The resulting density and velocity plots are very flat.  This is because in
Eq.~(\ref{M2d}) for $M^{2'}$ the denominator is very large with the
small value of $M^2$ while the numerator is small because of the factor $F$,
the slope, which is small in our model.  Therefore we have a very small change
in $M^2$ across the prominence and so $\bf V$ and $\rho$ change very little
as well.  Meanwhile there is more variation in the pressure and therefore the
temperature across the prominence.

\begin{figure*}
\begin{center}
\resizebox{0.49\hsize}{!}{\includegraphics*{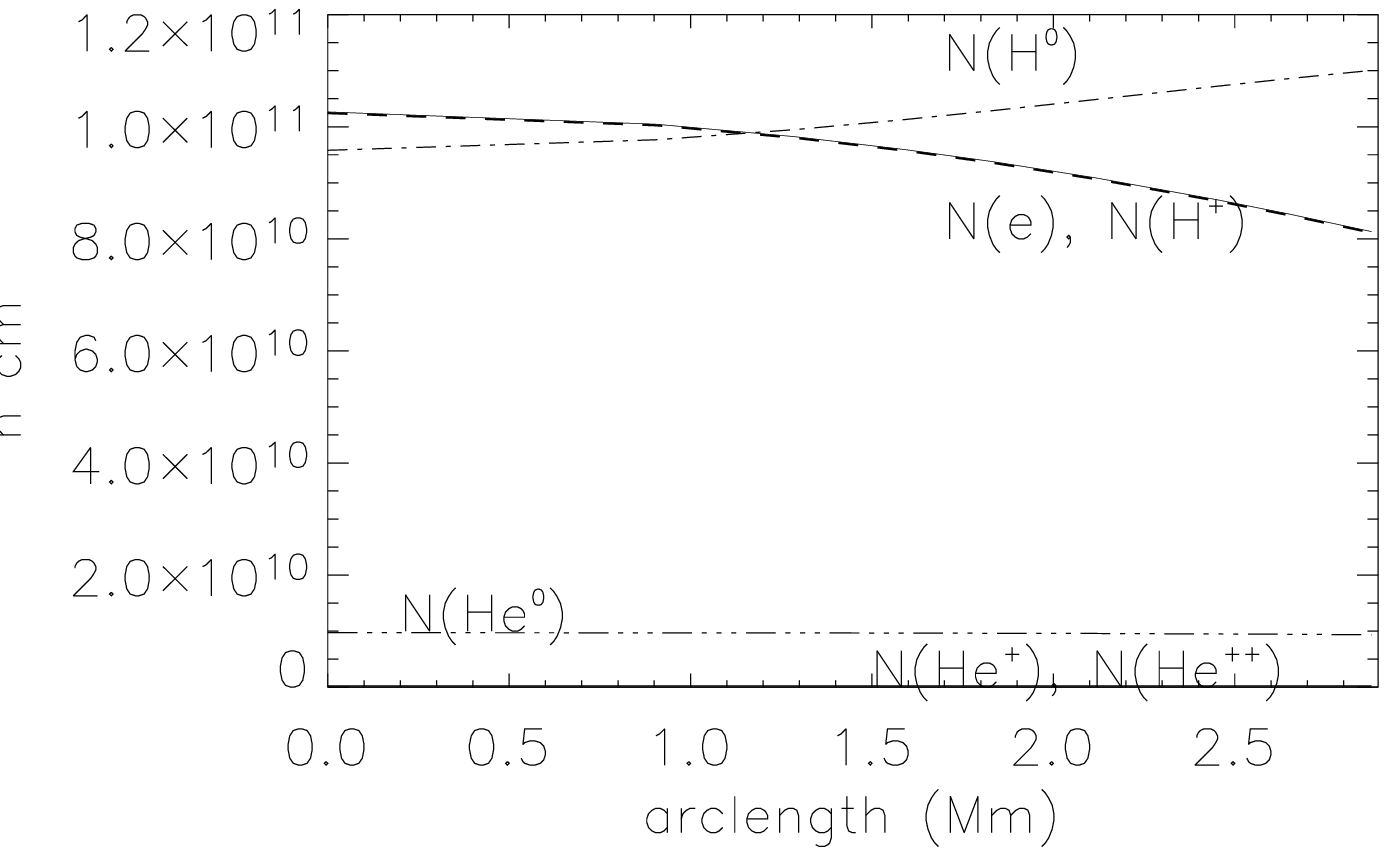}}
\resizebox{0.49\hsize}{!}{\includegraphics*{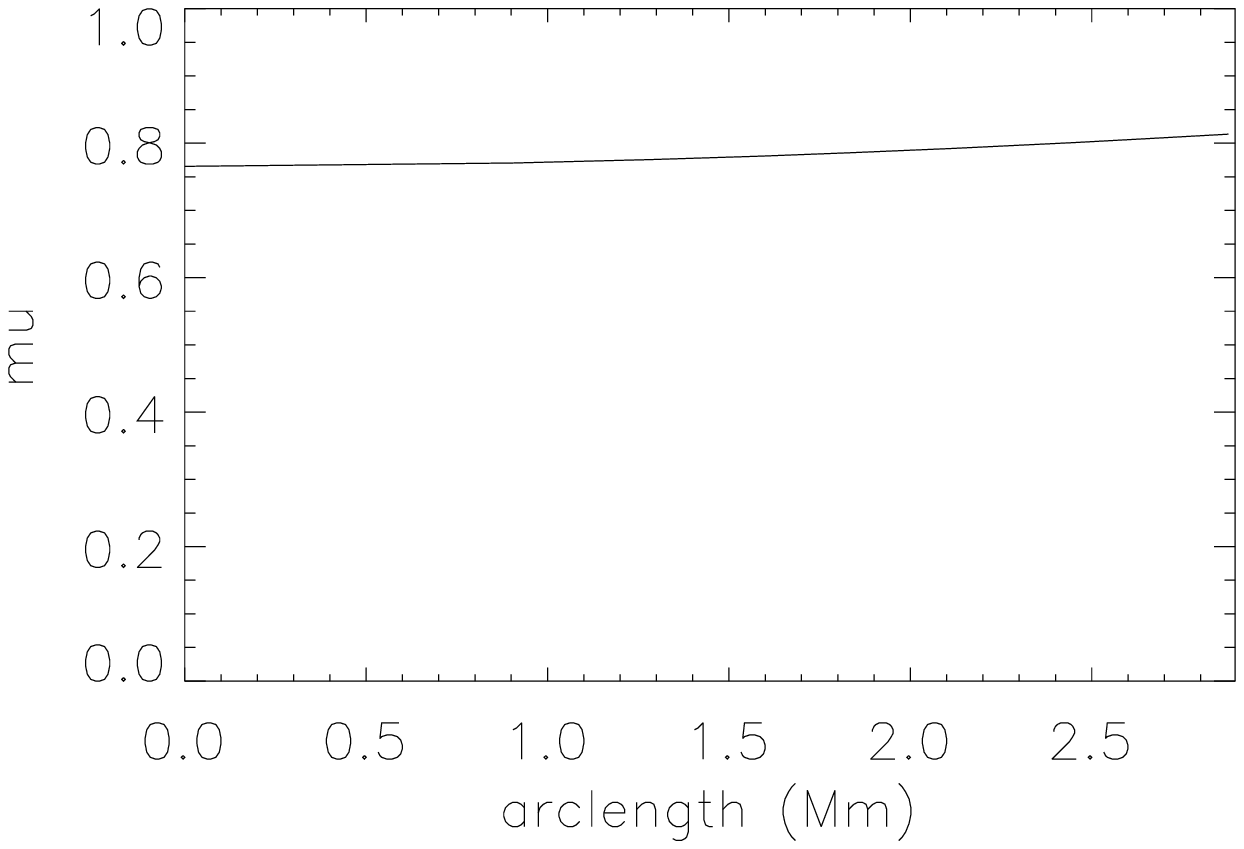}}
\end{center}
\caption{Abundances of the species $N(e)$, $N(H^0)$, $N(H^+)$, $N(He^0)$, $N(He^+)$ and $N(He^{++})$ (left picture) and mean particle mass scaled against the mass of a proton (right picture).  For a fully ionised hydrogen plasma the mean particle mass would be $0.5$ scaled against the mass of a proton.}
\label{abundances}
\end{figure*}

Besides protons and electrons, neutral hydrogen atoms, helium atoms and helium
ions ($He^+$ and $He^{++}$) are all present inside the prominence and so the
electron number density and gas density are not proportional to each other. 
The species populations are calculated for our model from Table~3 in Kuin \&
Poland~(1991).  While the helium ion populations are insignificant and the
helium atom population is a constant value around one tenth of the next
smallest population across the prominence, the hydrogen ionisation is visibly
greater at the hotter centre of the prominence than at the cooler edges
(Fig.~\ref{abundances}, left).  The proportion of gas density to electron
number density is larger at the edges than at the middle indicating that the
average particle mass has a maximum at the cool edge.  This is confirmed in
Fig.~\ref{abundances} (right).  The average particle mass $\mu$ is given in
terms of the mass of a proton $m_p$ by

\begin{equation}
\mu =m_p\frac{N(H^0)+N(H^+)+4N(He^0)+4N(He^+)+4N(He^{++})}{N(H^0)+2N(H^+)+N(He^0)+2N(He^+)+3N(He^{++})},
\label{mu}
\end{equation}

\noindent since $N(e)=N(H^+)+N(He^+)+2N(He^{++})$.  The average particle mass
inside the prominence ranges from around $0.75m_p$ in the middle of
the prominence to around $0.8m_p$ at the edges.  This compares to
$0.5m_p$ for a fully-ionised hydrogen plasma. 
Fig.~\ref{abundances} (left) shows that the maximum hydrogen ion/minimum
hydrogen atom populations occur at the middle and the minimum ion/maximum atom
populations at the edges.  The hydrogen ion (proton) and electron populations
are almost but not quite exactly equal across the prominence.  The hydrogen
ionisation ratio $N(H^+)/N(H^0)$ is approximately equal to $1$ across the
prominence, varying from about $1.25$ in the middle to about $0.8$ at the
edges.
\subsection{Applicability of ideal MHD}
\label{MHDapplicability}

Landman~(1983, 1984) calculated the line intensities to be expected under
non-LTE conditions and derived a value $n(H^+)/n(H^0)=0.07$.  Landman~(1986)
later revised his results upward by a factor of two.  Previous studies
obtained by Hirayama (see Hirayama,~1985) were an order of magnitude higher. 
Conditions in prominences are normally such that there are enough ions present
for the prominence to behave like a plasma and the ideal MHD equations are
valid.  For a prominence H=25,000 km high we calculate that under gravity the
plasma would free-fall to the photosphere in just over 7 mins.  Meanwhile
even with the neutral and ion number densities equal to each other
$n_n = n_e$, the magnetic diffusivity $\eta$ differs from the fully-ionised
case by less than one part in 1000 (Priest, 1982, p79): we
calculate $\eta =1272$~m$^2$/s taking the average particle mass to be the mass
of a proton$\times 2/3$ and the temperature to be 8000~K.  The time to diffuse
over 25,000~km is about 15,600~years.  Free-fall times and diffusion times
are comparable over distances of around 20~m.  This calculation is
sufficiently insensitive to ion:neutral ratios for our purposes.  Neutrals
would have to be many orders of magnitude more populous than ions for diffusion
to become important over the distances and timescales of interest.

Since the prominence plasma is only partially ionised and much denser than the ambient
coronal plasma, it is interesting to ask how well MHD describes the properties
of this prominence plasma. It is generally thought that prominences are supported
against the gravitational force by their magnetic field. Hence an especially important question is how 
well the frozen-in condition of ideal MHD is satisfied by the partially ionised
plasma. In particular, the neutral species (mainly H and He) are not directly supported
by the magnetic field but only through collisional coupling to the ions and electrons.
This leads to a drainage of the neutral species from the prominence.
The time scales for this drainage have recently been calculated by Gilbert et al. (2002)
using a simplified model involving the Lorentz force for the species with nonvanishing
electric charge, the gravitational force and collisional forces.
The species taken into account by the model are electrons, protons (H$^+$), singly charged
helium (He$^+$), neutral hydrogen (H) and neutral helium (He).
Based on these assumptions Gilbert et al.~(2002) find that for a prominence with a number
density
of $n=10^{10}$~cm$^{-3}$ and a vertical extension of $7000$ km ($0.01$ R$_\odot$) the
loss time scale for hydrogen is roughly $520$ hrs (22 days). The loss time for helium is
actually much shorter and of the order of 1 day.  For our example with atom
density around $10^{11}$~cm$^{-3}$ the Gilbert et al.~(2002) model gives
neutral helium and neutral hydrogen cross-field velocities of
$8.1\times 10^{-3}$ and $3.7\times 10^{-4}$ km/s respectively.  Therefore
over the time scales of interest this cross-field diffusion of neutrals is
negligible.  In one hour the helium atoms will drift about 20~km while the
hydrogen atoms, the larger category by far, would have travelled only about
1~km.  Ideal MHD with field-aligned flow is therefore a reasonable
description.

\begin{figure*}
\begin{center}
\resizebox{0.47\hsize}{!}{\includegraphics*{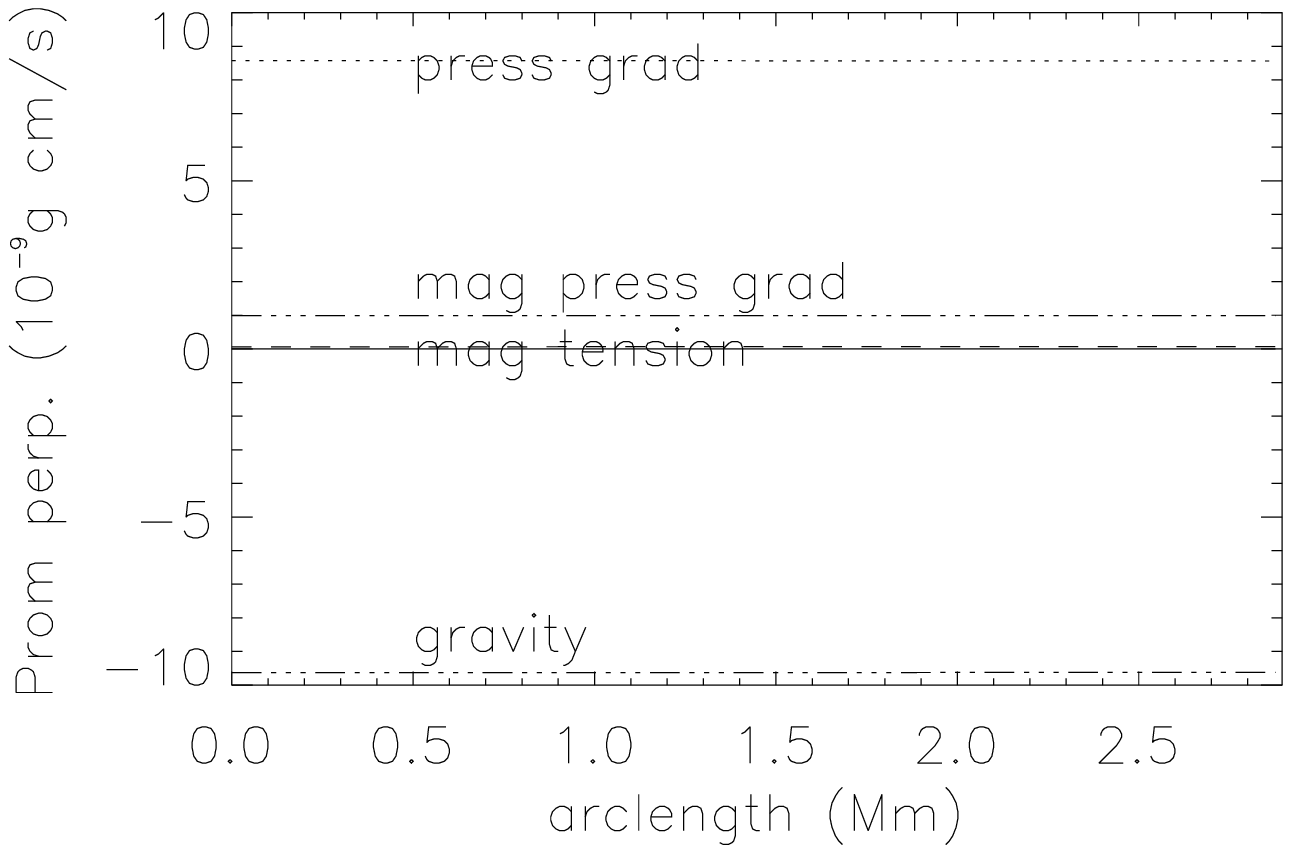}}
\resizebox{0.47\hsize}{!}{\includegraphics*{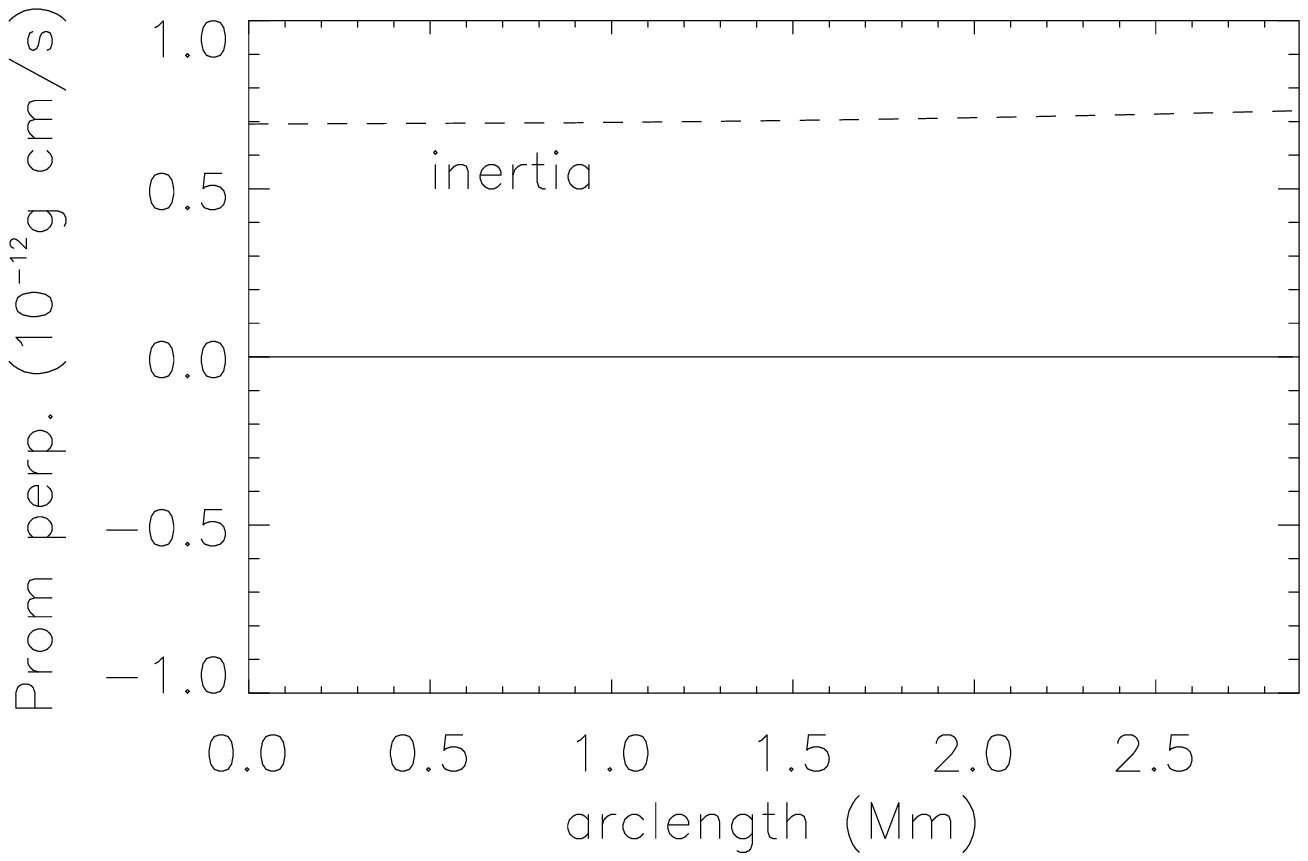}}
\resizebox{0.47\hsize}{!}{\includegraphics*{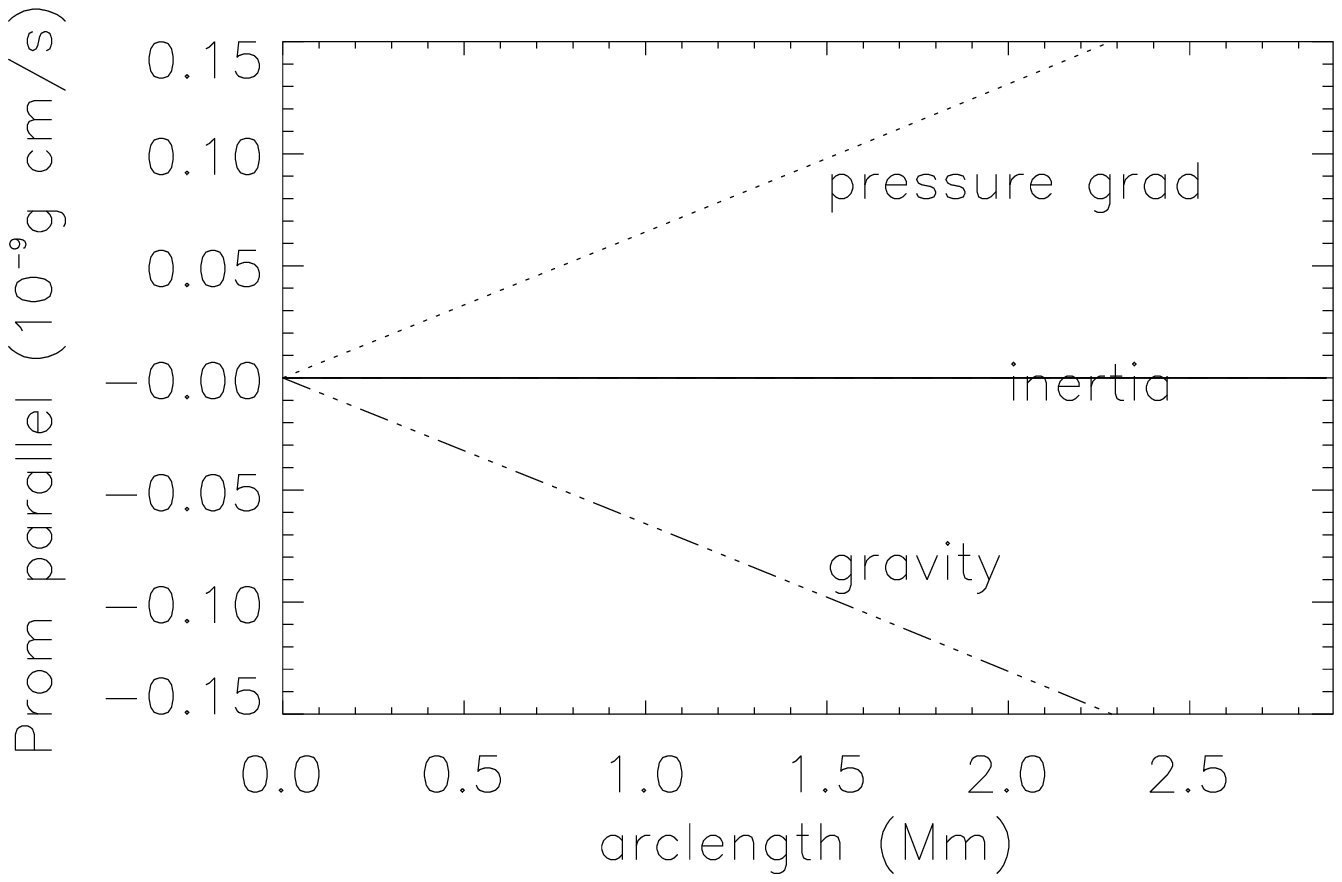}}
\resizebox{0.47\hsize}{!}{\includegraphics*{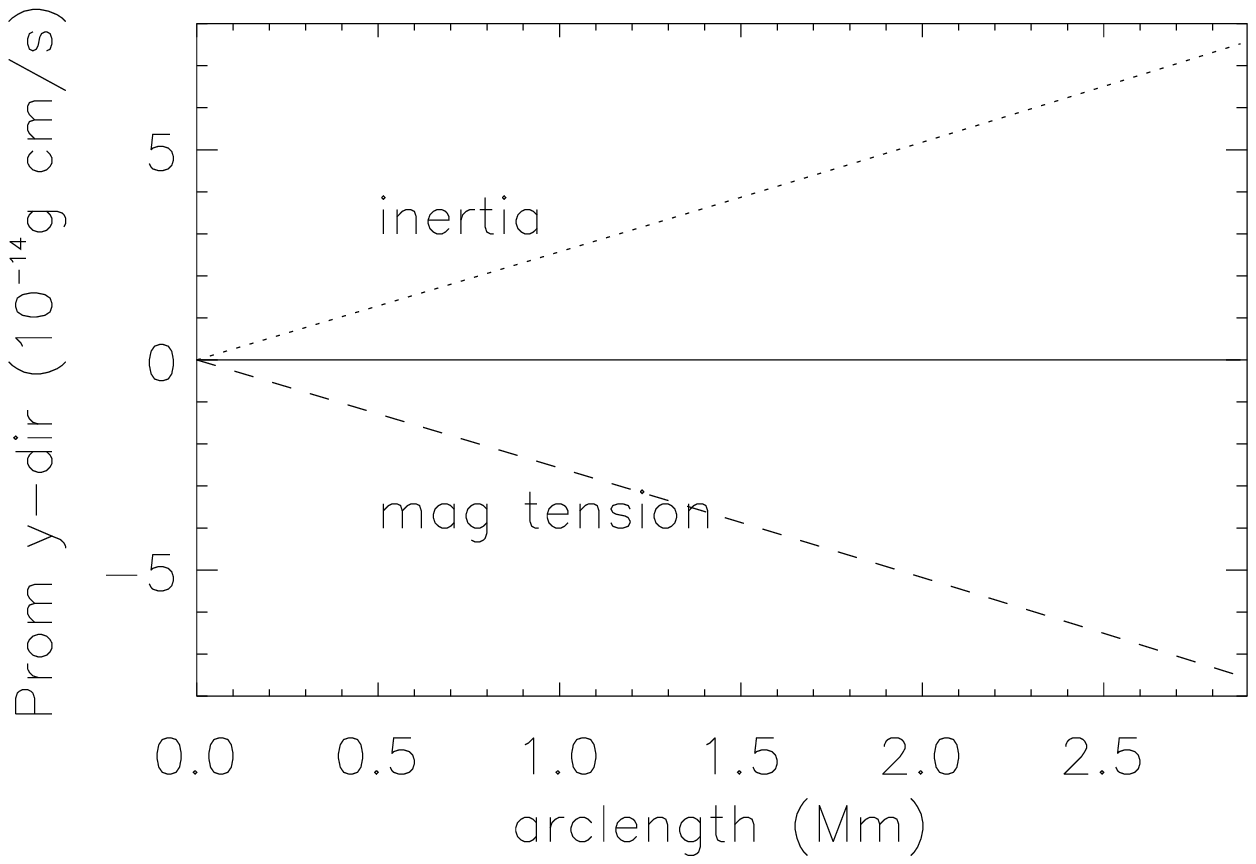}}
\end{center}
\caption{Shown is the momentum breakdown along and across the magnetic field. 
We graph components across the field line (top pictures), across the field line 
in the $X$-$Z$ plane (bottom left) and in the $Y$-direction (bottom right).  
}
\label{momen}
\end{figure*}

\subsection{Momentum balance across and along the prominence}
\label{momentumbalance}

Fig.~\ref{momen} shows the momentum breakdown in the present model.  
We graph components across the field line (top pictures), across the field 
line in the $x$-$z$ plane (bottom left) and in the $y$-direction (bottom right).
The field-aligned magnetic forces cancel because the Lorentz force is
perpendicular to the magnetic field line. 

Across the prominence field line the largest force is the downward
gravitational force because of the high density of the prominence plasma.  
The four remaining forces are all directed upward to balance this 
gravitational force. However, the gas pressure gradient force dominates all 
other forces to balance gravity. The magnetic tension force is much smaller
than the pressure gradient forces because the field line dip is very shallow. 
The inertial force is much smaller still because the flow velocity vector is
small, nearly straight and of a nearly constant size.  
Along the field lines the inertial force is negligible and the downward 
gravitational force is mostly balanced by an upward gas pressure force.
In the $Y$-direction the magnetic and gas pressures are zero because the model
is invariant in this direction.  Furthermore there is no gravitational force
component, so that the momentum balance in the $Y$-direction consists of the
inertial and magnetic tension forces.

\subsection{Energy balance across and along the prominence}
\label{energy}

\begin{figure*}
\begin{center}
\resizebox{0.49\hsize}{!}{\includegraphics*{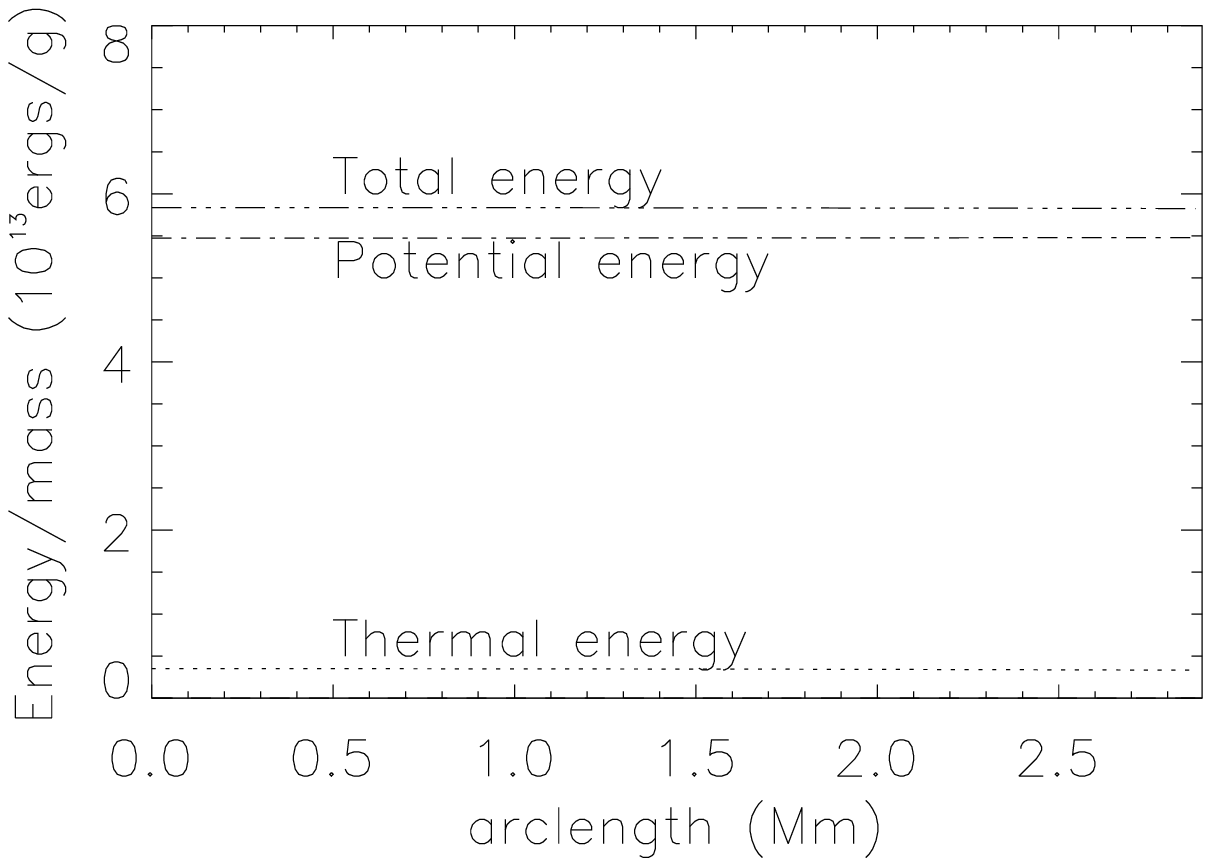}}
\resizebox{0.49\hsize}{!}{\includegraphics*{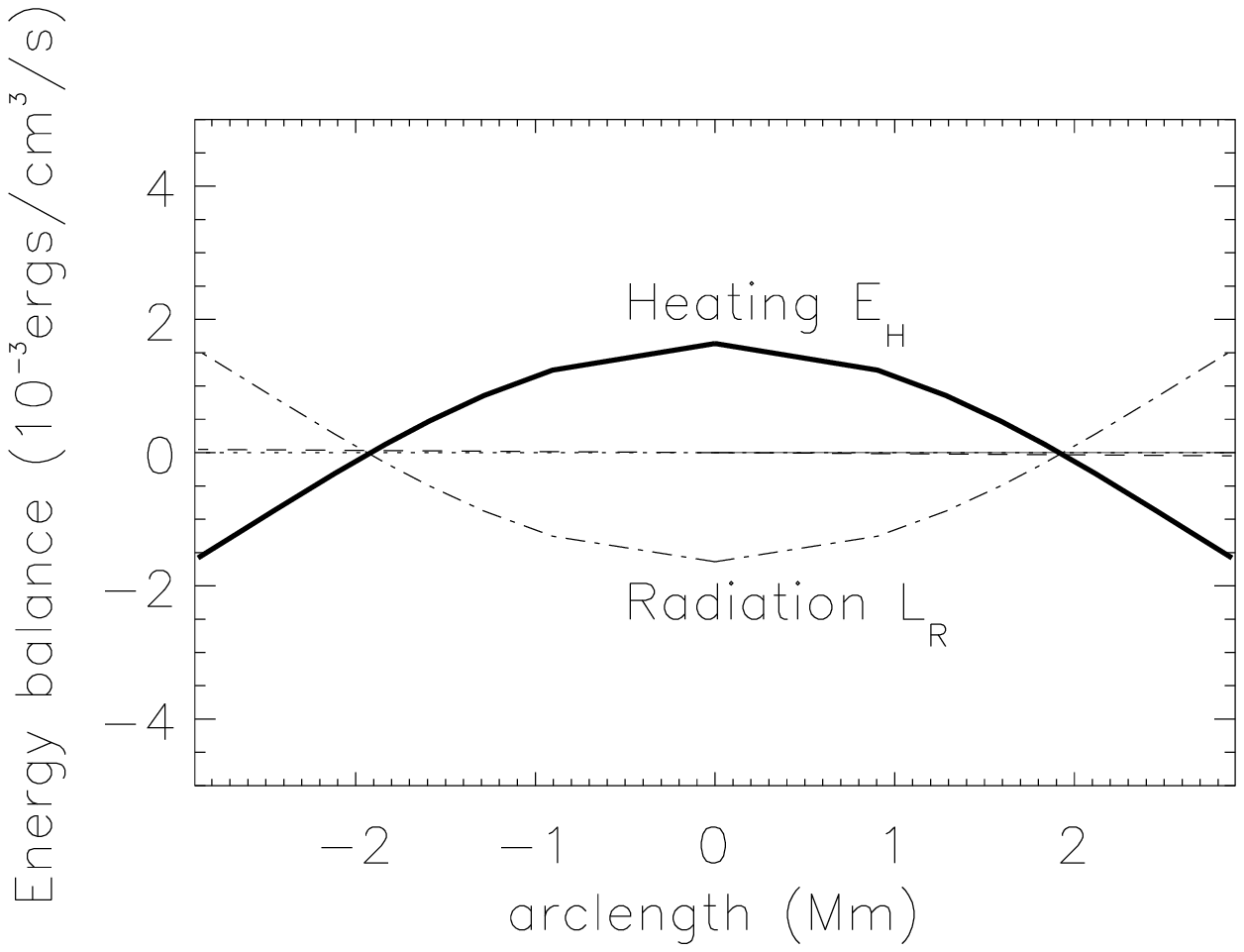}}
\end{center}
\caption{Energy (left picture) and heating (right picture) profiles of the prominence.  The potential energy is the dominannt energy of the prominence plasma because of the very large density of this plasma.  The thermal energy is much smaller while the kinetic energy is much smaller still.  The heating function $E_H$ is a nearly (but not quite) symmetric function of arclength.  This is because the radiative gains/losses $L_R$ (a symmetric function) are much larger than the conduction (a symmetric function) and the net heat in/out $q$ (an antisymmetric function), both too small to show in the graph.  It can be seen by the symmetry of $E_H$ that the flow has a negligible influence on the thermodynamics of the prominence plasma.}
\label{energyheat}
\end{figure*}

Fig.~\ref{energyheat} shows the breakdown of the total energy (left) and the
heating balance (right) inside the prominence.  The energy plots are graphed
only for the right half ($X>0$) of the prominence since the energy functions 
are all symmetric, while the heating balance is graphed for both halves of 
the prominence because there are asymmetries in the heating balance. 

Because of the much larger density of the prominence suspended at such a large 
height most of the energy contained in the prominence is potential energy.  The
thermal energy is smaller in the prominence than in the corona because of
the low temperature.  This can be seen by comparing Fig.~\ref{energyheat}(left) 
with the corresponding figures 5 and 8 (bottom left) of Paper 2 for the coronal 
loop model, i.e., in coronal loops the thermal energy dominates the total energy 
with only a small contribution from the potential energy.   

The heating balance on the other hand, contains the function $q$, the net heat 
in/out of the flow.  This net heat in/out of the flow $q$ is an anti-symmetric 
function, because from Eq. (3) $q$ is a derivative of a symmetric function (the 
antisymmetry of $q$ can be also seen in Paper 2, Figs. 5,8, bottom right pictures).  
Hence, if $q\neq 0$ this anti-symmetric function causes the heating function 
to be asymmetric as well.  The radiative loss and thermal conduction functions 
are both symmetric functions.  Thus, any asymmetry of the heating function can 
only be due to the flow, as in the case of the loop modelling of Paper 2.  
But the effect of the flow is negligible because of the small flow velocity of the 
model and thus deviations of the heating function from symmetry are practically 
negligible in the present model. 
 
The heating of the prominence is dominated to the exclusion of all else by a very large
radiative loss function.  Temperature gradients, which are present
because of the compressibility of the flow and the great density of the plasma
in the dip, cause some conduction to take place.  This conduction is
nevertheless small and is barely visible in the heating plot.  Although the
radiation dwarfs all else in the heat equation it is smaller than it would be
without non-LTE effects included.  In fact radiative gains from incident
coronal radiation are very significant.  The example plotted has flux tube 
width 500~km.  
Near the prominence edges there is a net radiative gain.  This radiative gain is due to incident radiation from the corona or chromosphere.  Even though an ambient coronal model is not explicitly included in our MHD model, effects of this incident radiation are taken into account in the radiation model of Kuin \& Poland~(1991) which we use here. 
 
Towards the middle there is a point where radiative gains and losses balance and in the middle, 
where the plasma is hotter because of plasma compression (see Fig. \ref{parampics}), there is a net radiative loss.  
Decreasing the flux tube width decreases the radiative losses.

We could have added to our model of a prominence dip a surrounding hot arcade model
separated from the prominence by a discontinuity as in Del Zanna \& Hood~(1996).  
However, we refrain from doing so for the following reason.  To satisfy
mass and momentum conservation across a discontinuity between a cool, dense
region and a hot, sparse region forces the enthalpy to change by a large
factor across the discontinuity.
Because the gravitational potential energy per unit mass is continuous
across the boundary and the influence of the kinetic energy is minimal on
each side for any realistic velocity, this change cannot be balanced.  Point
energy sources/sinks at such boundaries are
implied and energy changes by factors of 10 or more, corresponding to
delta-function behaviour in the heating function.

\section{Conclusions}
\label{conclusions}

We have modelled a prominence by using a two-dimensional compressible equilibrium
solution of the full ideal steady MHD equations with a consistent heating included 
in the model for the first time.  Our model generalises known self-similar prominence
models, such as those by 
Hood \& Anzer~(1990), 
and Del Zanna \& Hood~(1996). 
The heating model takes into account non-LTE radiation for the first time in
a full MHD model by exploiting a radiative transfer model by Kuin \&
Poland~(1991).  Although the model is 2D, a third component of the magnetic and
velocity vector fields allow us to model the highly sheared fields observed in
prominences.  Unlike the coronal loop model in Paper 2, the heat in/out of
the flow does not influence the energy equation significantly.  The model is
consistent with an ionisation ratio of order unity according to the radiation
model of Kuin \& Poland~(1991).  This is consistent with several observations.
Both magnetic diffusion and cross-field diffusion of neutrals are found to be
insignificant within the time scales of interest so that an ideal MHD
description is reasonable.  The modelled prominence dip must be very shallow
for the physical parameters to stay within reasonable bounds. This is also consistent with observations.  Within the prominence the plasma is so dense that the
gas pressure bears most of the burden of the prominence weight.  The supporting role
of the magnetic field may be more important underneath the prominence where
the plasma is less dense and the magnetic field may be compressed, and
therefore stronger and more dipped, than inside the prominence.  We were unable to add self-consistently a surrounding hotter arcade solution separated from the 
cooler prominence either with an MHD discontinuity, because this would imply a huge 
enthalpy change there, or with a tangential discontinuity between thermally isolated prominence and coronal field lines. Such a global prominence model remains a challenge for 
the future.    

\begin{acknowledgements}
We thank BC Low for useful comments about MHD shocks.  The authors
acknowledge support from the EC's human potential programme under contract
HPRN-CT-2000-00153, PLATON.
\end{acknowledgements}

\end{document}